\documentclass[11pt,preprint]{elsarticle}

\usepackage{amsmath,amssymb,amsthm,paralist,hyperref}
\usepackage{tikz}

\newtheorem{lemma}{Lemma}

\newcommand{\cG}{\mathcal G}
\renewcommand{\phi}{\varphi}
\renewcommand{\epsilon}{\varepsilon}

\newcommand{\gr}{\mathrm{gr}}
\newcommand{\rel}{\mathrm{rel}}
\newcommand{\out}{\mathrm{out}}
\newcommand{\inn}{\mathrm{in}}
\newcommand{\enc}{\mathsf{enc}}
\newcommand{\midd}{\mathrm{mid}}

\newlength\ys

\journal{Information Processing Letters}

\title{\bf Computability by Monadic Second-Order Logic}

\author{Joost Engelfriet}
\ead{j.engelfriet@liacs.leidenuniv.nl}
\address{LIACS, Leiden University, P.O.~Box~9512, \\
  2300~RA~Leiden, the Netherlands}

\begin{document}

\begin{abstract} 
\noindent
A binary relation on graphs is recursively enumerable if and only if it can be computed by 
a formula of monadic second-order logic. The latter means that the formula defines a set of graphs, 
in the usual way, such that each ``computation graph'' in that set determines a pair 
consisting of an input graph and an output graph. 
\end{abstract}


\maketitle 

\noindent
There are many characterizations of computability,
but the one presented here does not seem to appear explicitly in the literature.\footnote{This 
first sentence and the first part of the next sentence are taken over from~\cite{eng07}.
} 
Nevertheless, it is a natural and simple characterization,
based on the intuitive idea that a computation of a machine, or a derivation of a grammar,
can be represented by a graph satisfying a formula of monadic second-order (MSO) logic. 
Assuming the reader to be familiar with MSO logic on graphs (see, e.g., \cite[Chapter~5]{coueng12}),
the MSO-computability of a binary relation on graphs can be given in half a page, see below.
One advantage of the definition is that there is no need to code the graphs as strings or numbers.

For an alphabet $\Psi$, we consider directed edge-labeled graphs $g=(V,E)$ over~$\Psi$ where $V$ is a nonempty finite set of nodes and $E\subseteq V\times \Psi\times V$ is a set of labeled edges. 
We also denote $V$ by $V_g$, and $E$ by $E_g$. An edge $(u,\psi,v)\in E_g$ is called a $\psi$-edge.
Isomorphic graphs are considered to be equal. 
The set of all (abstract) graphs over $\Psi$ is denoted by $\cG_\Psi$.

To model computations we use a special edge label $\nu$ that is not in $\Psi$. 
We define a \emph{computation graph} over $\Psi$ to be a graph $h$ over $\Psi\cup\{\nu\}$ 
with at least one $\nu$-edge such that 
for every $u,v,u',v'\in V_h$,
\begin{compactitem}
\item[(1)] $(u,\nu,u)\notin E_h$, and
\item[(2)] if $(u,\nu,v),(u',\nu,v') \in E_h$, then $(u,\nu,v') \in E_h$.
\end{compactitem}
The \emph{input graph} $\inn(h)$ is defined to be the subgraph of $h$ induced by all nodes that have an outgoing $\nu$-edge, and the \emph{output graph} $\out(h)$ is the subgraph of~$h$ induced by all nodes that have an incoming $\nu$-edge. By~(2) above, the $\nu$-edges of $h$ connect every node of $\inn(h)$ to every node of $\out(h)$, and so by~(1) above, $V_{\inn(h)}$ and $V_{\out(h)}$ are disjoint. 
In fact, the role of the $\nu$-edges is just to specify 
an ordered pair of disjoint subsets of $V_h$, in a simple way. 
Note that there may be arbitrarily many nodes and edges in $h$ that belong neither to $\inn(h)$ nor to $\out(h)$.
Also, there may be edges between $\inn(h)$ and $\out(h)$ other than the $\nu$-edges. 
This notion of computation graph generalizes the ``pair graph'' of~\cite{engvog20}, which on its turn generalizes the ``origin graph'' of~\cite{bojdavguipen17}.

For a set $H$ of computation graphs over $\Psi$ we define the \emph{graph relation computed by $H$} to be
$\rel(H)=\{(\inn(h),\out(h))\mid h\in H\}\subseteq \cG_\Psi\times \cG_\Psi$.
Finally, for an alphabet $\Gamma$, we say that 
a graph relation $R\subseteq \cG_\Gamma\times\cG_\Gamma$ is \emph{MSO-computable}
if there are an alphabet $\Delta$ and an MSO-definable set $H$ of computation graphs over $\Gamma\cup\Delta$
such that $\rel(H)=R$. As observed before, 
we assume the reader to be familiar with MSO logic on graphs.\footnote{The atomic 
formulas of MSO logic are $x=y$, $x\in X$, and $\mathrm{edge}_\psi(x,y)$, where $x$ and~$y$ are nodes,
$X$ is a set of nodes, and $\mathrm{edge}_\psi(x,y)$ expresses 
that there is a $\psi$-edge from $x$ to~$y$.
}
The closed MSO formula $\phi$ that defines the set $H$ can be viewed as a ``machine'' 
of which the computations are represented by the graphs in $H$. 
We will also say that $\rel(H)$ is the graph relation computed by $\phi$.
For each \mbox{$h\in H$}, the input graph $\inn(h)$ and the output graph $\out(h)$ must be 
graphs over the input/output alphabet $\Gamma$. 
The \emph{auxiliary alphabet} $\Delta$ is needed to allow the edges of a computation graph 
that are not part of its input or output graph, to carry arbitrary information in their label; 
it is similar to the ``working alphabet'' of a machine. 
This notion of MSO-computability generalizes the \mbox{``MSO-expressibility''} of graph relations 
of~\cite{engvog20},\footnote{The relation $R$ is ``MSO-expressible'', 
in the sense of~\cite[Section~3.1]{engvog20}, if it is MSO-computable by a set $H$ of pair graphs,
where a pair graph is a computation graph $h$ such that $V_h=V_{\inn(h)}\cup V_{\out(h)}$.
}
which on its turn generalizes the MSO graph transductions 
of~\cite[Chapter~7]{coueng12} (as shown in~\cite[Section~7.1]{engvog20}).

\begin{figure}[t]
  \begin{center}
\begin{tikzpicture}
\begin{scope}[xscale=0.6,yscale=0.6]
%
\draw [fill] (2,0) circle [radius=0.07];
\draw [fill] (0,2) circle [radius=0.07];
\draw [fill] (4,2) circle [radius=0.07];
\draw [fill] (2,4) circle [radius=0.07];
\draw [->] (2,0) -- (1,1); \draw (1,1) -- (0,2);
\draw [->] (0,2) -- (1,3); \draw (1,3) -- (2,4);
\draw [->] (2,0) -- (3,1); \draw (3,1) -- (4,2);
\draw [->] (4,2) -- (3,3); \draw (3,3) -- (2,4);
\draw [->] (2,4) -- (2,2); \draw (2,2) -- (2,0);
\node at (0.65,0.65) {$\gamma$}; 
\node at (0.65,3.15) {$\gamma$};
\node at (3.35,0.65) {$\beta$};
\node at (3.45,3.15) {$\beta$};
\node at (1.55,2) {$\alpha$};
\draw [fill] (12,0) circle [radius=0.07];
\draw [fill] (10,2) circle [radius=0.07];
\draw [fill] (12,4) circle [radius=0.07];
\draw [->] (12,0) -- (11,1); \draw (11,1) -- (10,2);
\draw [->] (10,2) -- (11,3); \draw (11,3) -- (12,4);
\draw [->] (12,4) -- (12,2); \draw (12,2) -- (12,0);
\draw [->] (12,0) -- (7,0); \draw (7,0) -- (2,0);
\draw [->] (10,2) -- (7,2); \draw (7,2) -- (4,2);
\draw [->] (12,4) -- (7,4); \draw (7,4) -- (2,4);
\node at (10.65,0.65) {$\beta$}; 
\node at (10.65,3.15) {$\beta$};
\node at (12.45,2) {$\alpha$};
\node at (7,0.45) {$d$};
\node at (7,2.45) {$d$};
\node at (7,4.45) {$d$};
\draw (2,2) ellipse (3.5 and 3);
\draw (11,2) ellipse (2.5 and 3);
\draw (4,-0.5) .. controls (5,-1.3) and (8.6,-1.3) .. (9.6,-0.5);
\draw [->] (6.99,-1.1) -- (7,-1.1);
\node at (7,-0.75) {$\nu$};
\end{scope}
\end{tikzpicture}
    \end{center}
  \caption{A computation graph $h$ for an induced subgraph, with $\Gamma=\{\alpha,\beta,\gamma\}$. The input graph $\inn(h)$ and output graph $\out(h)$ are surrounded by ovals, and the $\nu$-labeled edge from the left oval to the right oval represents the 12 $\nu$-labeled edges from each node of $\inn(h)$ to each node of $\out(h)$.}
   \label{fig:compgraph}
  \end{figure}
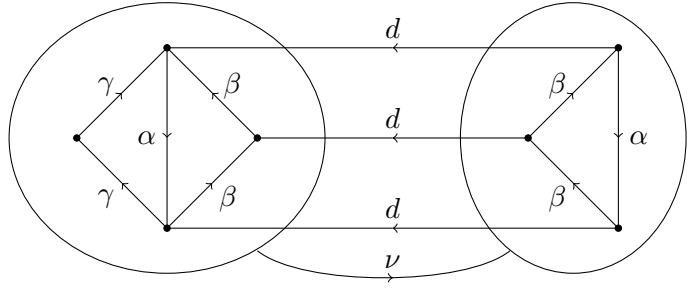

\medskip
\noindent
\textbf{Examples.}
(1) Let $R\subseteq \cG_\Gamma\times\cG_\Gamma$ be the set of all $(g,g')$ 
such that $g'$ is an induced subgraph of $g$. The graph relation $R$ is MSO-computable because it can be computed by an MSO-definable set $H$ of computation graphs over $\Gamma\cup\Delta$, with $\Delta=\{d\}$.
We note that, by definition, the set of all computation graphs~$h$ over $\Gamma\cup\Delta$ is MSO-definable,
and the sets of nodes $V_{\inn(h)}$ and $V_{\out(h)}$ can be expressed in MSO logic. 
The set $H$ consists of computation graphs $h$ such that $V_h=V_{\inn(h)}\cup V_{\out(h)}$, $\inn(h)$ and $\out(h)$ are graphs over $\Gamma$, and the $d$-edges form an isomorphism from $\out(h)$ to an induced subgraph of $\inn(h)$. The last condition means, in detail, that for every $u,v,u',v'\in V_h$,
\begin{compactitem}
\item if $(u,d,v)$ is an edge of $h$, then $u\in V_{\out(h)}$ and $v\in V_{\inn(h)}$,
\item if $u\in V_{\out(h)}$, then $u$ has an outgoing $d$-edge,
\item if $(u,d,v)$ and $(u',d,v')$ are edges of $h$, then 
    \begin{compactitem}
    \item $u=u'$ if and only if $v=v'$, and
    \item for every $\gamma\in\Gamma$, $(u,\gamma,u')\in E_h$ if and only if $(v,\gamma,v')\in E_h$.
    \end{compactitem}
\end{compactitem}
There may be $\gamma$-edges in $h$ between $\inn(h)$ and $\out(h)$, with $\gamma\in\Gamma$;
though they are harmless, we could additionally forbid them. 
For an example of such a computation graph see Fig.~\ref{fig:compgraph}.
Obviously the above conditions can be expressed by an MSO formula $\phi$, which defines~$H$. 
Moreover $\rel(H)=R$, and hence $R$ is MSO-computable. 
Note that $R$ is even ``MSO-expressible'', in the sense of~\cite{engvog20}. 

As another (similar) example, if $R$ consists of all $(g,g')$ such that $g$ has at least two, disjoint, 
induced subgraphs isomorphic to $g'$, then we take $\Delta=\{d_1,d_2\}$, we require that the $d_i$-edges
satisfy the same conditions as the $d$-edges above (for each $i\in\{1,2\}$), and we require that no node of $\inn(h)$ has both an incoming $d_1$-edge and an incoming $d_2$-edge. 

(2) Let $g_0$ be a fixed graph over $\Gamma$, and let $R\subseteq \cG_\Gamma\times\cG_\Gamma$ be 
the set of all $(g,g_0)$ such that the number of nodes of $g$ with an outgoing $\alpha$-edge equals 
its number of nodes with an outgoing $\beta$-edge, with $\alpha,\beta\in\Gamma$. 
There is an MSO-definable set $H$ of computation graphs over $\Gamma\cup\Delta$ such that $\rel(H)=R$,
where $\Delta=\{d,e\}$. It consists of all graphs $h$ that are obtained 
by adding $\nu$-, $d$- and $e$-edges to the disjoint union of $g$, $g'$, and $g_0$, 
where $g$ is an arbitrary graph over $\Gamma$ and $g'$ is isomorphic to $g$. 
The $\nu$-edges determine that $\inn(h)=g$ and $\out(h)=g_0$. 
The $d$-edges establish an isomorphism between $g$ and $g'$, and 
the $e$-edges establish a bijection between the nodes of $g$ with an outgoing $\alpha$-edge 
and the nodes of $g'$ with an outgoing $\beta$-edge. 
Since these requirements can easily be expressed in MSO logic, $R$ is MSO-computable. 
It is not difficult to show that $R$ is not ``MSO-expressible'', cf. the Conclusion of~\cite{engvog20}.
\qed

\medskip
Our aim is now to prove the following theorem.

\medskip
\noindent
\textbf{Theorem.}
A graph relation is MSO-computable if and only 
if it is recursively enumerable. 

\begin{figure}[t]
  \begin{center}
\begin{tikzpicture}
\begin{scope}[xscale=0.8,yscale=0.6]
%
\draw [fill] (0,6) circle [radius=0.07];
\draw [fill] (2,6) circle [radius=0.07];
\draw [fill] (4,6) circle [radius=0.07];
\draw [fill] (6,6) circle [radius=0.07];
\draw [fill] (8,6) circle [radius=0.07];
\draw [fill] (0,4) circle [radius=0.07];
\draw [fill] (2,4) circle [radius=0.07];
\draw [fill] (4,4) circle [radius=0.07];
\draw [fill] (6,4) circle [radius=0.07];
\draw [fill] (8,4) circle [radius=0.07];
\draw [fill] (0,2) circle [radius=0.07];
\draw [fill] (2,2) circle [radius=0.07];
\draw [fill] (4,2) circle [radius=0.07];
\draw [fill] (6,2) circle [radius=0.07];
\draw [fill] (8,2) circle [radius=0.07];
\draw [fill] (0,0) circle [radius=0.07];
\draw [fill] (2,0) circle [radius=0.07];
\draw [fill] (4,0) circle [radius=0.07];
\draw [fill] (6,0) circle [radius=0.07];
\draw [fill] (8,0) circle [radius=0.07];
\draw [->] (0,6) -- (1,6); \draw (1,6) -- (2,6); \node at (1,6.35) {$i$};
\draw [->] (2,6) -- (3,6); \draw (3,6) -- (4,6); \node at (3,6.35) {$\alpha$};
\draw [->] (4,6) -- (5,6); \draw (5,6) -- (6,6); \node at (5,6.35) {$\beta$};
\draw [->] (6,6) -- (7,6); \draw (7,6) -- (8,6); \node at (7,6.35) {$B$};
\draw [->] (0,6) -- (0,5); \draw (0,5) -- (0,4);
\draw [->] (2,6) -- (2,5); \draw (2,5) -- (2,4);
\draw [->] (4,6) -- (4,5); \draw (4,5) -- (4,4);
\draw [->] (6,6) -- (6,5); \draw (6,5) -- (6,4);
\draw [->] (8,6) -- (8,5); \draw (8,5) -- (8,4);
\draw [->] (0,4) -- (1,4); \draw (1,4) -- (2,4); \node at (1,4.35) {$\beta$};
\draw [->] (2,4) -- (3,4); \draw (3,4) -- (4,4); \node at (3,4.35) {$i$};
\draw [->] (4,4) -- (5,4); \draw (5,4) -- (6,4); \node at (5,4.35) {$\beta$};
\draw [->] (6,4) -- (7,4); \draw (7,4) -- (8,4); \node at (7,4.35) {$B$};
\draw [->] (0,4) -- (0,3); \draw (0,3) -- (0,2);
\draw [->] (2,4) -- (2,3); \draw (2,3) -- (2,2);
\draw [->] (4,4) -- (4,3); \draw (4,3) -- (4,2);
\draw [->] (6,4) -- (6,3); \draw (6,3) -- (6,2);
\draw [->] (8,4) -- (8,3); \draw (8,3) -- (8,2);
\draw [->] (0,2) -- (1,2); \draw (1,2) -- (2,2); \node at (1,2.35) {$\beta$};
\draw [->] (2,2) -- (3,2); \draw (3,2) -- (4,2); \node at (3,2.35) {$\alpha$};
\draw [->] (4,2) -- (5,2); \draw (5,2) -- (6,2); \node at (5,2.35) {$i$};
\draw [->] (6,2) -- (7,2); \draw (7,2) -- (8,2); \node at (7,2.35) {$B$};
\draw [->] (0,2) -- (0,1); \draw (0,1) -- (0,0);
\draw [->] (2,2) -- (2,1); \draw (2,1) -- (2,0);
\draw [->] (4,2) -- (4,1); \draw (4,1) -- (4,0);
\draw [->] (6,2) -- (6,1); \draw (6,1) -- (6,0);
\draw [->] (8,2) -- (8,1); \draw (8,1) -- (8,0);
\draw [->] (0,0) -- (1,0); \draw (1,0) -- (2,0); \node at (1,0.35) {$\beta$};
\draw [->] (2,0) -- (3,0); \draw (3,0) -- (4,0); \node at (3,0.35) {$\alpha$};
\draw [->] (4,0) -- (5,0); \draw (5,0) -- (6,0); \node at (5.1,0.4) {$f$};
\draw [->] (6,0) -- (7,0); \draw (7,0) -- (8,0); \node at (7,0.35) {$B$};
\draw (4,6) ellipse (2.5 and 0.8);
\draw (2,0) ellipse (2.5 and 0.8);
\draw (2,6.5) .. controls (-2,8) and (-1,4) .. (-0.25,0.35);
\draw [->] (-0.901,4.01) -- (-0.9,4);
\node at (-0.63,4) {$\nu$};
\end{scope}
\end{tikzpicture}
    \end{center}
  \caption{A computation graph $h$ over $\Gamma\cup\Delta$, with $\Gamma=\{\alpha,\beta\}$ and $\Delta=\{i,f,B,*\}$. It represents the computation of a (very simple) Turing machine $M$ that changes every $\alpha$ of the input string into $\beta$ and vice versa. Here the input string is $\alpha\beta$, and $M$ uses space $m=3$ and time $n=4$. The initial state of $M$ is $i$, the final state is $f$, and the instructions are $i\alpha\vdash \beta i$, $i\beta\vdash \alpha i$, and $iB\vdash fB$, where $B$ is the blank. The strings $w_1,w_2,w_3,w_4$ corresponding to $M$'s computation are $i\alpha\beta B$, $\beta i\beta B$, $\beta\alpha iB$, and $\beta\alpha fB$. 
The $*$-labels of the vertical edges of $h$ are omitted.} 
  \label{fig:turing} 
  \end{figure}
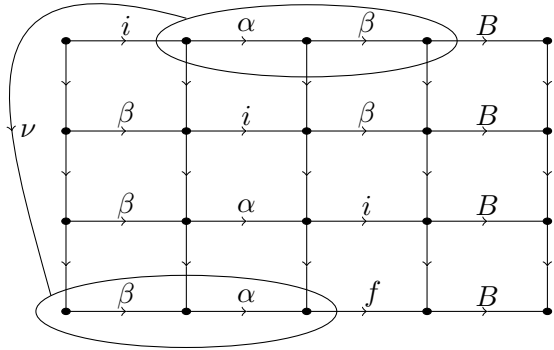

\medskip
Recursive enumerability of a graph relation $R$  means that 
there is a (single tape) nondeterministic Turing machine $M$ 
such that $(g,g')\in R$ if and only if, on input $g$,
$M$ has a computation that outputs $g'$. In one direction this theorem is obvious: every MSO-computable graph relation is recursively enumerable. In fact, on input $g\in\cG_\Gamma$ 
(coded as a string in an appropriate way) 
$M$ guesses a computation graph $h$ over $\Gamma\cup\Delta$ such that $\inn(h)=g$, checks whether 
$h$ satisfies the MSO formula $\phi$ (cf.~\cite[Chapter~6]{coueng12}), and if so, 
outputs the (coded) graph $\out(h)$. 
To show the other direction we first consider the case of string relations. 
For the notion of MSO-computability we represent a string $w=\gamma_1\gamma_2\cdots \gamma_k$ over $\Gamma$ 
by the graph $\gr(g)\in\cG_\Gamma$ such that 
$V_{\gr(g)}=\{1,2,\dots,k+1\}$ and $E_{\gr(g)} = \{(j,\gamma_j,j+1)\mid 1\leq j\leq k\}$. 
The proof is similar to the one of~\cite[Theorem~5.6]{coueng12}.
Let $M$ be a nondeterministic Turing machine that computes 
the recursively enumerable string relation $R\subseteq \Gamma^*\times\Gamma^*$. 
Consider a computation of~$M$ that, for an input string $w$, outputs the string $w'$. 
Suppose that it uses space $m$ and time $n$. 
Thus, it can be viewed as a sequence of strings $w_1,\dots,w_n$, each of length $m+1$,
such that $w_i$ is the content of $M$'s tape at time~$i$ (including the state of $M$),
$w_1$ contains $w$ (plus the initial state and blanks), 
and $w_n$ contains $w'$ (and a final state and blanks). Clearly, this sequence can be 
represented by a grid of dimension $n\times (m+2)$. 
The rows of the grid are the graphs $\gr(w_1),\dots,\gr(w_n)$, 
which are connected by $*$-labeled column edges from the $j$-th node of $w_i$ to the $j$-th node of $w_{i+1}$ for every $1\leq i\leq n-1$ and $1\leq j\leq m+2$.
It is easy to turn that grid into a computation graph~$h$ by adding $\nu$-edges from the nodes of $\gr(w)$ in the first row to those of $\gr(w')$ in the last row. 
Thus, $h$ is a computation graph over $\Gamma\cup\Delta$ such that $\inn(h)=\gr(w)$ and $\out(h)=\gr(w')$, where the alphabet~$\Delta$ consists of the column symbol~$*$, the working symbols of $M$ (including the blank), and the states of $M$. 
For an example see Fig.~\ref{fig:turing}.
Since the set of grids is MSO-definable (as shown in~\cite[Section~5.2]{coueng12}), it is a straightforward exercise in MSO logic to show that the computation graphs~$h$, obtained from the (successful) computations of $M$, can be defined by an MSO formula~$\phi_M$. In particular, $\phi_M$ should express that the consecutive rows of the grid (corresponding to strings $w_i$ and $w_{i+1}$) satisfy the (local) changes determined by the instructions of $M$. This shows that the graph relation computed by $\phi_M$ is $\gr(R)=\{(\gr(w),\gr(w'))\mid (w,w')\in R\}$, and so, $\gr(R)$ is MSO-computable. 

For an alphabet $\Gamma$, let the \emph{graph encoding relation} $\enc_\Gamma$ consist of all pairs $(g,\gr(w))$ such that $g\in\cG_\Gamma$ and $w$ is an appropriate encoding of $g$ as a string
(which we will specify later).\footnote{Appropriateness means that the encoding and the corresponding decoding are computable in a straightforward intuitive sense. In particular, it is decidable whether or not a given string is the encoding of some graph. Any standard encoding of graphs satisfies these requirements. 
}  
By definition, if a graph relation $R\subseteq \cG_\Gamma\times\cG_\Gamma$ is recursively enumerable then 
there is a recursively enumerable string relation $R'$ such that 
$R$ is the composition of $\enc_\Gamma$, $\gr(R')$, and $\enc_\Gamma^{-1}$. 
Hence, to obtain our theorem for graph relations it now suffices to prove the following two lemmas.

\begin{lemma}\label{comp}\rm
The class of MSO-computable graph relations is closed under inverse and composition.
\end{lemma}

\begin{lemma}\label{enc}\rm
For every $\Gamma$, the graph encoding relation $\enc_\Gamma$ is MSO-computable. 
\end{lemma}

\textit{Proof of Lemma~\ref{comp}.}
Closure under inverse is obvious: just reverse the direction of all $\nu$-edges. 
To prove closure under composition, let $R_1$ and $R_2$ be graph relations 
computed by MSO formulas $\phi_1$ and $\phi_2$. We may assume that $\phi_1$ and $\phi_2$
use the same auxiliary alphabet $\Delta$. Moreover, we may assume that 
every computation graph $h$ defined by $\phi_1$ or $\phi_2$ is connected: if not, then 
add a special symbol $\mu$ to $\Delta$ and require that every node $u$ of~$h$ 
that is not in $\inn(h)$ or $\out(h)$, has a $\mu$-edge to $\inn(h)$ or $\out(h)$. 
Finally, we assume that $\phi_1$ uses the label $\nu_1$ instead of $\nu$, and 
$\phi_2$ uses $\nu_2$ instead of $\nu$, with $\nu_1\neq\nu_2$. 
The MSO formula~$\phi$ that computes the composition of $R_1$ and $R_2$, uses the auxiliary alphabet 
$\Delta\cup\{\nu_1,\nu_2,d\}$ and defines computation graphs $h$ that are obtained 
as the disjoint union of a computation graph $h_1$ of $\phi_1$ and a computation graph 
$h_2$ of $\phi_2$, enriched by $d$-edges that establish an isomorphism between $\out(h_1)$ and $\inn(h_2)$, 
and by $\nu$-edges from $\inn(h_1)$ to $\out(h_2)$. It should be clear that this can be realized
by $\phi$; for instance, it expresses that the connected components of $h$ 
minus its enriching edges satisfy $\phi_1$ or $\phi_2$,
depending on whether they contain a $\nu_1$-edge or a \mbox{$\nu_2$-edge}. 
\qed

\medskip
\textit{Proof of Lemma~\ref{enc}.}
We first specify the relation~$\enc_\Gamma$. 
Let $g\in \cG_\Gamma$. 
We may assume that $V_g$ is the set of strings $\{a,a^2,\dots,a^n\}$ 
over the alphabet~$\{a\}$, for some $n\geq 1$, where $a\notin\Gamma$. 
Let $E_g=\{(u_1,\gamma_1,v_1),\dots,(u_m,\gamma_m,v_m)\}$ 
for some $m\geq 0$. We encode $g$, in a standard way, as the string 
\[
w = \#a\#a^2\# \cdots \#a^n \$u_1\gamma_1v_1\$ \cdots \$u_m\gamma_mv_m\$
\]
over the alphabet $\Omega=\Gamma\cup\{a,\#,\$\}$, 
and we define the graph encoding relation $\enc_\Gamma\subseteq \cG_\Gamma\times \cG_\Omega$ 
to consist of all pairs $(g,\gr(w))$. 
Note that since $w$ depends on linear orderings of $V_g$ and $E_g$, a graph $g$ has in general more than one encoding. On the other hand, the relation $\enc^{-1}_\Gamma$ is a function.
\begin{figure}[t]
  \begin{center}
\begin{tikzpicture}
\begin{scope}[xscale=0.6,yscale=0.6]
%
\draw [fill] (2,5) circle [radius=0.07];
\draw [fill] (4,5) circle [radius=0.07];
\draw [fill] (6,5) circle [radius=0.07];
\draw [fill] (8,5) circle [radius=0.07];
\draw [fill] (10,5) circle [radius=0.07];
\draw [fill] (12,5) circle [radius=0.07];
\draw [fill] (14,5) circle [radius=0.07];
\draw [fill] (16,5) circle [radius=0.07];
\draw [fill] (18,5) circle [radius=0.07];
\draw [fill] (20,5) circle [radius=0.07];
\draw [fill] (0,2) circle [radius=0.07];
\draw [fill] (2,2) circle [radius=0.07];
\draw [fill] (4,2) circle [radius=0.07];
\draw [fill] (6,2) circle [radius=0.07];
\draw [fill] (8,2) circle [radius=0.07];
\draw [fill] (10,2) circle [radius=0.07];
\draw [fill] (12,2) circle [radius=0.07];
\draw [fill] (14,2) circle [radius=0.07];
\draw [fill] (16,2) circle [radius=0.07];
\draw [fill] (18,2) circle [radius=0.07];
\draw [->] (2,5) -- (3,5); 
\draw [->] (3,5) -- (5,5);
\draw [->] (5,5) -- (7,5);
\draw [->] (7,5) -- (9,5); 
\draw [->] (9,5) -- (11,5);
\draw [->] (11,5) -- (13,5);
\draw [->] (13,5) -- (15,5); 
\draw [->] (15,5) -- (17,5);
\draw [->] (17,5) -- (19,5);
\draw (19,5) -- (20,5);
\draw [->] (0,2) -- (1,2);
\draw [->] (1,2) -- (3,2); 
\draw [->] (3,2) -- (5,2);
\draw [->] (5,2) -- (7,2);
\draw [->] (7,2) -- (9,2); 
\draw [->] (9,2) -- (11,2);
\draw [->] (11,2) -- (13,2);
\draw [->] (13,2) -- (15,2); 
\draw [->] (15,2) -- (17,2);
\draw (17,2) -- (18,2);
\node at (3,5.45) {$\#$};
\node at (5,5.45) {$a$};
\node at (7,5.45) {$\#$};
\node at (9,5.45) {$a$};
\node at (11,5.45) {$a$};
\node at (13,5.45) {$\#$};
\node at (15,5.45) {$a$};
\node at (17,5.45) {$a$};
\node at (19,5.45) {$a$};
\node at (1,1.65) {$a$};
\node at (3,1.65) {$\gamma$};
\node at (5,1.65) {$a$};
\node at (7,1.65) {$\$$};
\node at (9,1.65) {$a$};
\node at (11,1.65) {$\gamma$};
\node at (13,1.65) {$a$};
\node at (15,1.65) {$a$};
\node at (17,1.65) {$\$$};
\draw [->] (0,2) -- (2.6,4); \draw (2.6,4) -- (4,5);
\draw [->] (2,2) -- (4.6,4); \draw (4.6,4) -- (6,5);
\draw [->] (4,2) -- (4,4); \draw (4,4) -- (4,5);
\draw [->] (6,2) -- (6,4); \draw (6,4) -- (6,5);
\draw [->] (8,2) -- (5.4,4); \draw (5.4,4) -- (4,5);
\draw [->] (10,2) -- (7.4,4); \draw (7.4,4) -- (6,5);
\draw [->] (12,2) -- (9.4,4); \draw (9.4,4) -- (8,5);
\draw [->] (14,2) -- (11.4,4); \draw (11.4,4) -- (10,5);
\draw [->] (16,2) -- (13.4,4); \draw (13.4,4) -- (12,5);
\draw (4,5) .. controls (5,7) and (7,7) .. (8,5); \draw [->] (5.99,6.5) -- (6,6.5);
\draw (6,5) .. controls (7,7) and (9,7) .. (10,5); \draw [->] (7.99,6.5) -- (8,6.5);
\draw (8,5) .. controls (9.5,7) and (12.5,7) .. (14,5); \draw [->] (10.99,6.5) -- (11,6.5);
\draw (10,5) .. controls (11.5,7) and (14.5,7) .. (16,5); \draw [->] (12.99,6.5) -- (13,6.5);
\draw (12,5) .. controls (13.5,7) and (16.5,7) .. (18,5); \draw [->] (14.99,6.5) -- (15,6.5);
\node at (6,6.85) {$\alpha$};
\node at (8,6.85) {$\alpha$};
\node at (11,6.85) {$\alpha$};
\node at (13,6.85) {$\alpha$};
\node at (15,6.85) {$\alpha$};
\node at (1,3.25) {$\delta$};
\node at (3,3.25) {$\delta$};
\node at (4.3,3.25) {$\delta$};
\node at (5.7,3.25) {$\delta$};
\node at (7,3.25) {$\delta$};
\node at (9,3.25) {$\delta$};
\node at (11,3.25) {$\delta$};
\node at (13,3.25) {$\delta$};
\node at (15,3.25) {$\delta$};
\draw (20,5) .. controls (21,-2) and (0,0) .. (0,2);
\draw [->] (9.01,0.1) -- (9,0.1);
\node at (9,0.55) {$\$$};
\end{scope}
\end{tikzpicture}
    \end{center}
  \caption{\label{fig:grplus} The graph $\gr^+(w)$ for the string $w=\#a\#aa\#aaa\$a\gamma a\$a\gamma aa\$$, which is an encoding of the graph $g$ with $V_g=\{a,aa,aaa\}$ and $E_g=\{(a,\gamma,a),(a,\gamma,aa)\}$. The graph $\gr(w)$ is obtained from $\gr^+(w)$ by removing all $\alpha$- and $\delta$-edges. }
  \end{figure}
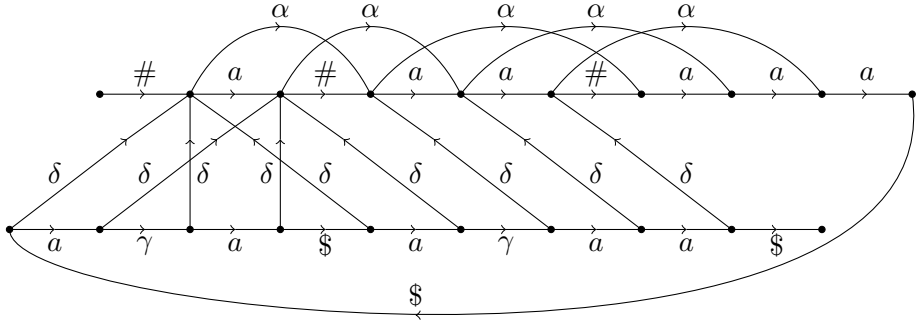
The set of strings over $\Omega$ that encode graphs over $\Gamma$ is not a regular language, and hence 
the set $\enc_\Gamma(\cG_\Gamma)$ of graphs over $\Omega$ is not MSO-definable~\cite{buc60,elg61,tra62}.
However, by enriching each $\gr(w)$ with $\alpha$-edges and $\delta$-edges 
(where $\alpha$ and $\delta$ are special symbols not in $\Omega$),
we can turn $\enc_\Gamma(\cG_\Gamma)$ into an MSO-definable set of graphs. 
For a string $w$ as displayed above we define $\gr^+(w)$ to be the graph $\gr(w)$
to which $\alpha$-edges and $\delta$-edges are added as follows. 
For an example see Fig.~\ref{fig:grplus}.
The $\alpha$-edges allow an MSO formula to express the fact that 
the first half of $w$ is of the form $\#a\#a^2\# \cdots \#a^n \$$.
For each substring $\#a^i\#a^i$ of $w$ ($1\leq i\leq n-1$) there are $\alpha$-edges in $\gr^+(w)$ 
from the nodes of the first occurrence of $\gr(a^i)$ in $\gr(w)$ to the nodes of the second occurrence 
of $\gr(a^i)$ in $\gr(w)$, such that they form an isomorphism between these two subgraphs.
An MSO formula on $\gr^+(w)$ can express that $w$ is in the regular language
$\#a(\#a^*)^*(\$a^*\Gamma a^*)^*\$ $, 
and, using the outgoing $\alpha$-edges of $\gr(\#a^i\#)$,
it can enforce that each substring $\#a^i\#$ is followed by $a^{i+1}\#$ or $a^{i+1}\$$.
The $\delta$-edges in $\gr^+(w)$ witness the fact that for each substring $\$u_j\gamma_jv_j\$$
of $w$ ($1\leq j\leq m$) both $u_j$ and $v_j$ are in $\{a,a^2,\dots,a^n\}$, 
i.e., $u_j$ and $v_j$ are ``declared'' in the first half of $w$. 
Thus, there are $\delta$-edges from the nodes of $\gr(u_j)$ to the nodes of some $\gr(\#a^i\#)$ or 
$\gr(\#a^i\$)$ in the first half of $\gr(w)$
that establish an isomorphism between $\gr(u_j)$ and $\gr(a^i)$, and similarly for $\gr(v_j)$.
This can also easily be expressed by an MSO formula. 
Moreover, the $\delta$-edges can be used to express that an edge is not encoded twice in $w$, i.e., 
if $j\neq k$ then $\$u_j\gamma_jv_j\$ \neq \$u_k\gamma_kv_k\$$; 
in fact, $u_j=u_k$ if and only if the two $\delta$-edges that start from the first nodes of $\gr(u_j)$ and $\gr(u_k)$ in $\gr^+(w)$, lead to the same node (and similarly for $v_j=v_k$).
We now define $\enc^+_\Gamma$ to consist of all pairs $(g,\gr^+(w))$ where $w$ encodes $g$.
It follows that the set $\enc^+_\Gamma(\cG_\Gamma)$ is MSO-definable.\footnote{We recall
that the set of graphs $\gr(w)$, where $w$ is an arbitrary string over $\Omega$, is MSO-definable, see for instance~\cite[Corollary~5.12]{coueng12} or~\cite[Example~2.1]{engvog20}.
} 

Finally, we show that $\enc_\Gamma\subseteq \cG_\Omega\times \cG_\Omega$ is MSO-computable by describing 
the computation graphs $h$ over $\Omega\cup\Delta$ 
in an MSO-definable set $H$ such that $\rel(H)=\enc_\Gamma$. 
The auxiliary alphabet is $\Delta=\{\alpha,\delta,d,e\}$. 
Let $\midd(h)$ be the subgraph of $h$ induced by the nodes of $h$ that are not incident with a $\nu$-edge,
i.e., that are not in $V_{\inn(h)}$ or $V_{\out(h)}$. 
First, we require that $\midd(h)$ is in $\enc^+_\Gamma(\cG_\Gamma)$, i.e., $\midd(h)=\gr^+(w)$ 
where $w$ encodes some graph $g$ in $\cG_\Gamma$. 
Second, we require that there are $d$-edges from $\out(h)$ to $\midd(h)$ that establish 
an isomorphism between $\out(h)$ and the graph obtained from $\midd(h)$ 
by removing all $\alpha$- and $\delta$-edges. This means that $\out(h)=\gr(w)$. 
Third, it remains to require that $\inn(h)$ is isomorphic to $g$. To realize this, we require that 
$\inn(h)\in\cG_\Gamma$ and that there are $e$-edges from $\inn(h)$ to $\midd(h)$ that establish a bijection 
between $V_{\inn(h)}$ and the nodes of $\midd(h)$ that have an incoming \mbox{$\#$-edge} 
(thus representing a bijection between $V_{\inn(h)}$ and $V_g=\{a,a^2,\dots,a^n\}$). 
Since we wish this bijection to represent an isomorphism between $\inn(h)$ and $g$, 
we require for every $(x,\gamma,y)\in V_{\inn(h)}\times\Gamma\times V_{\inn(h)}$
that $(x,\gamma,y)$ is an edge of $\inn(h)$ if and only if 
there exist nodes $x',x'',y',y''$ of $\midd(h)$ such that 
\begin{compactitem}
\item[(1)]$(x,e,x')$ and $(y,e,y')$ are edges of $h$, 
\item[(2)] $(x'',\delta,x')$ and $(y'',\delta,y')$ are edges of $\midd(h)$,
\item[(3)] $x''$ has an incoming $\$$-edge in $\midd(h)$, and 
\item[(4)] there is a directed path from $x''$ to $y''$ in $\midd(h)$, 
of which the consecutive edge labels form a string in $a^*\gamma$.
\end{compactitem}
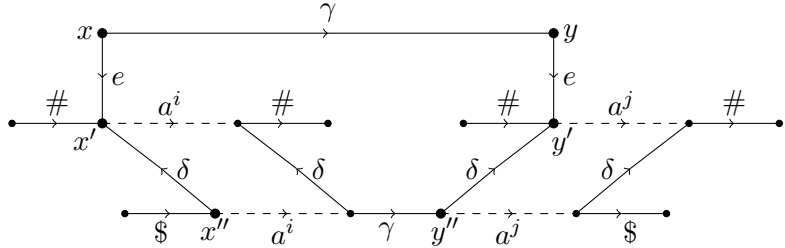
\begin{figure}[t]
  \begin{center}
\begin{tikzpicture}
\begin{scope}[xscale=0.6,yscale=0.6]
%
\draw [fill] (2,5) circle [radius=0.1];
\node at (1.62,5) {$x$};
\draw [fill] (12,5) circle [radius=0.1];
\node at (12.38,5) {$y$};
\draw [->] (2,5) -- (7,5); \draw (7,5) -- (12,5);
\node at (7,5.45) {$\gamma$};
\draw [fill] (0,3) circle [radius=0.07];
\draw [fill] (2,3) circle [radius=0.1];
\node at (1.62,2.65) {$x'$};
\draw [fill] (5,3) circle [radius=0.07];
\draw [fill] (7,3) circle [radius=0.07];
\draw [fill] (10,3) circle [radius=0.07];
\draw [fill] (12,3) circle [radius=0.1];
\node at (12.2,2.6) {$y'$};
\draw [fill] (15,3) circle [radius=0.07];
\draw [fill] (17,3) circle [radius=0.07];
\draw [->] (0,3) -- (1,3); \draw (1,3) -- (2,3);
\node at (1,3.45) {$\#$};
\draw [dashed,->] (2,3) -- (3.6,3); \draw [dashed] (3.5,3) -- (5,3);
\node at (3.5,3.45) {$a^i$};
\draw [->] (5,3) -- (6,3); \draw (6,3) -- (7,3);
\node at (6,3.45) {$\#$};
\draw [->] (10,3) -- (11,3); \draw (11,3) -- (12,3);
\node at (11,3.45) {$\#$};
\draw [dashed,->] (12,3) -- (13.6,3); \draw [dashed] (13.5,3) -- (15,3);
\node at (13.5,3.45) {$a^j$};
\draw [->] (15,3) -- (16,3); \draw (16,3) -- (17,3);
\node at (16,3.45) {$\#$};
\draw [fill] (2.5,1) circle [radius=0.07];
\draw [fill] (4.5,1) circle [radius=0.1];
\node at (4.5,0.65) {$x''$};
\draw [fill] (7.5,1) circle [radius=0.07];
\draw [fill] (9.5,1) circle [radius=0.1];
\node at (9.6,0.6) {$y''$};
\draw [fill] (12.5,1) circle [radius=0.07];
\draw [fill] (14.5,1) circle [radius=0.07];
\draw [->] (2.5,1) -- (3.5,1); \draw (3.5,1) -- (4.5,1);
\node at (3.3,0.6) {$\$$};
\draw [dashed,->] (4.5,1) -- (6.1,1); \draw [dashed] (6,1) -- (7.5,1);
\node at (6,0.60) {$a^i$};
\draw [->] (7.5,1) -- (8.5,1); \draw (8.5,1) -- (9.5,1);
\node at (8.3,0.6) {$\gamma$};
\draw [dashed,->] (9.5,1) -- (11.1,1); \draw [dashed] (11,1) -- (12.5,1);
\node at (11,0.6) {$a^j$};
\draw [->] (12.5,1) -- (13.5,1); \draw (13.5,1) -- (14.5,1);
\node at (13.7,0.6) {$\$$};
\draw [->] (2,5) -- (2,4); \draw (2,4) -- (2,3);
\node at (2.35,4) {$e$};
\draw [->] (12,5) -- (12,4); \draw (12,4) -- (12,3);
\node at (12.35,4) {$e$};
\draw [->] (4.5,1) -- (3.3,2); \draw (3.3,2) -- (2,3);
\node at (3.8,2) {$\delta$};
\draw [->] (7.5,1) -- (6.3,2); \draw (6.3,2) -- (5,3);
\node at (6.8,2) {$\delta$};

\draw [->] (9.5,1) -- (10.7,2); \draw (10.7,2) -- (12,3);
\node at (10.2,2) {$\delta$};
\draw [->] (12.5,1) -- (13.7,2); \draw (13.7,2) -- (15,3);
\node at (13.2,2) {$\delta$};
\end{scope}
\end{tikzpicture}
    \end{center}
  \caption{\label{fig:enc} Parts of a computation graph $h$ showing the MSO-computability of $\enc_\Gamma$. The nodes $x$ and $y$ belong to $\inn(h)$, all other nodes to $\midd(h)$.}
  \end{figure}
This situation is sketched in Fig.~\ref{fig:enc}.
Condition~(1) means that $x$ and $y$ correspond to substrings $\#a^i*$ and $\#a^j*$ of~$w$ 
(with $*\in\{\#,\$\}$), i.e., to nodes $a^i$ and $a^j$ of $g$,
and conditions (2)-(4) mean that $w$ has a substring $\$a^i\gamma a^j\$$, i.e., 
that $(a^i,\gamma,a^j)$ is an edge of $g$.  
It should be clear that all these requirements can be expressed in MSO logic,
and that the graph relation computed by $H$ is $\enc_\Gamma$. 
\qed

\medskip
Lemma~\ref{enc} is trivial from the point of view of Turing computability: if $w$ encodes $g$, then both $g$ and $\gr(w)$ can be represented by $w$ on the tape of a Turing machine. This is however based on the intuition that our encoding of graphs as strings is computable. Since the notion of MSO-computability discussed here uses graphs as datatype rather than strings, we were able to give a formal proof of that intuition. The reader may object that the formal proof is based on the intuition that the encoding of a string $w$ as the graph $\gr(w)$ is computable. One might then argue that the latter encoding is simpler than the former. 

Traditionally, it has been shown that MSO logic is related to regularity, e.g., 
to regular string languages~\cite{buc60,elg61,tra62} and
regular tree languages~\cite{don70,thawri68}.
If one identifies regularity with computability by a finite-state machine, 
then this approach fails for MSO logic on graphs, because 
``no notion of finite graph automaton has been defined 
that would generalize conveniently finite automata on words and terms'' 
(\cite[Section~1.7]{coueng12}).
For this reason, the MSO transducers of~\cite[Chapter~7]{coueng12} were proposed 
to play the role of finite-state transducers of graphs,
and in the case of strings they indeed turned out to be equivalent 
to two-way finite-state transducers~\cite{enghoo01}.
We have shown above how, dropping the finite-state condition, 
MSO logic is related to computability by any machine. 

If, on the other hand, one identifies regularity with rationality, i.e., 
with a smallest class containing all finite sets of objects and closed under a number of natural operations
on sets of objects (union, concatenation, and Kleene star in the case of string languages), 
then the class of all MSO-definable sets of graphs has a rational characterization~\cite{eng91}.
Since the recursively enumerable string relations also have a rational characterization
(as discussed in~\cite{eng07}), 
the question remains whether there is a natural rational characterization 
of the MSO-computable graph relations. Such a characterization would at least involve the operations of 
union, composition, and transitive closure of graph relations.  

The above quote from~\cite[Section~1.7]{coueng12} refers to the non-existence of a finite-state graph automaton 
that accepts exactly the MSO-definable sets of graphs. In~\cite{tho91} 
a finite-state graph acceptor is introduced of which the computations are ``tilings'' of the input graphs 
(which have to be graphs of bounded degree).
All ``tiling-recognizable'' sets of graphs accepted by these machines are MSO-definable, 
and the reverse is true for strings and trees.
If we would allow the nodes of our graphs to have labels, then we could model the input graph $\inn(h)$ and 
the output graph $\out(h)$ of a computation graph~$h$ by two special node labels rather than by $\nu$-edges. 
Then, similar to MSO-computability, we could define a graph relation to be ``tiling-computable'' by requiring 
the set $H$ of computation graphs to be tiling-recognizable rather than MSO-definable. 
This leads to the following question for graphs of bounded degree: 
is every recursively enumerable graph relation tiling-computable? 
Note that, as shown in~\cite[Example~3.2(b)]{tho91}, the set of grids is tiling-recognizable.

Descriptive complexity theory investigates logics that characterize complexity classes. 
By Fagin's theorem (see, e.g., \cite[Theorem~5.1]{fag93}), 
the complexity class NP equals the set of problems that can be specified by
existential second-order formulas. In terms of graphs, such a formula requires the existence of 
an extension of the input graph by additional labeled hyperedges (where a hyperedge is a sequence of nodes), 
such that the resulting (hyper)graph satisfies a first-order formula. In our notion of MSO-computability 
we require that the input graph is an induced subgraph of a graph 
that satisfies a monadic second-order formula, and we obtain all recursively enumerable problems.  

We finally note that the notion of MSO-computability can easily be generalized to deal with arbitrary 
relational structures (cf.~\cite[Section~5.1]{coueng12}). 

\medskip
\textbf{Acknowledgement.} I thank the reviewers for their helpful suggestions.

\medskip
\noindent
\textbf{That's all folks!}
This was my last paper. Thank you, dear reader, and farewell.

\end{document}